\documentstyle[12pt,epsf]{article}
\begin{document}

\begin{center}
{\LARGE \bf NUCLEAR THEORY 1995}
\bigskip

{\bf Report on the Town Meeting held at Argonne National Laboratory}

{\bf January 29-30,1995}
\bigskip

Conveners: George Bertsch, Berndt M\"uller, John Negele

Advisors: James Friar, Vijay Pandharipande
\medskip

{\sc Final Version}
\end{center}

\bigskip

\section{Summary}

\subsection{Overview}

The nuclear theory town meeting was attended by close to 100 participants
representing the broad range of research in nuclear theory in the United
States.  The large attendance, which was more than twice that of the 1989
town meeting, demonstrates the liveliness and enthusiasm pervading the
field of nuclear theory during the past five years.
Thirteen speakers were invited to review progress
in different areas of nuclear theory, and there were special reports
from the other town meetings, from representatives of the funding
agencies, and from the Institute of Nuclear Theory.  The complete
program of the town meeting is listed in the Appendix.
This document makes an attempt to collect some of the important advances
in nuclear theory during the past five years and identifies challenges
and opportunities for future research.

Important progress has been made since the time of the last Long Range
Plan in all major areas of theoretical nuclear physics. Advances in
computational techniques have greatly enhanced our ability to address
nuclear structure problems from the microscopic point of view. The
Schr\"odinger equation has been solved for the ground and excited
states of few-nucleon systems, yielding unique information on the
nuclear force. The improved microscopic understanding of nuclear structure
has important applications to electroweak interactions in nuclei and
nuclear astrophysics, as well as to the interpretation of electron
scattering data. Lattice calculations and QCD vacuum models have
provided crucial insight into the quark structure of the nucleon
and other hadrons. Advances in nuclear transport theory has made
new tools available for the quest of extracting the nuclear equation
of state from intermediate and high energy heavy ion collisions, as
well as for quantitative predictions of the initial conditions to be
reached in nuclear collisions at RHIC.

The increased complexity of nuclear interactions at the higher energies
reached at CEBAF, and in the future at RHIC, poses unprecedented challenges
for nuclear theory. A thorough understanding of the implications of the
data that will emerge from these two facilities will require massive
theoretical efforts. Theorists from many different areas of nuclear
physics will have to work together, in collaboration with experimentalists,
to identify meaningful observables for new physics, to reliably
calculate background effects, and to solve the underlying nuclear and hadron
structure physics. Nuclei far off the island of stability
will provide stringent new tests for theories of nuclear structure,
and a significant theoretical effort will be needed to derive the
astrophysical implications of their observed properties. A fundamental,
quantitative understanding of the structure and interactions
of hadrons in terms of quantum chromodynamics remains an essential
long-term goal of nuclear theory.

\subsection{Recommendations}

In order to meet these challenges and progress towards our long-range goals,
we make the following recommendations:

\begin{enumerate}
{\bf
\item The strength of nuclear theory groups at both universities and
national laboratories must be commensurate with the challenges posed
by the ongoing and future experimental initiatives. The creation of
new positions for young theorists should be encouraged, e.g., with
bridging arrangements, joint appointments, and distinguished
fellowship programs.

\item We support the continuation of the Institute for Nuclear Theory,
and give high priority to its successful programs and workshops focussed
on forefront areas of importance to both theory and experiment.

\item In order to be able to address important scientific problems and
to maintain their leading role in computational science, nuclear theorists
must have access to state-of-the-art high performance computing platforms.
Adequate local computing facilities for research groups must also be
assured.}
\end{enumerate}

\subsection{Highlights}

Here we list a small number of recent achievements that highlight the
scope of progress in nuclear theory since the 1989 Long Range Plan. These
topics are to meant provide salient examples for the breadth of progress
in theoretical nuclear physics and are not the only important achievements.
More detail on the highlighted topics as well as other progress is
provided in subsequent sections.

\subsubsection{Nuclear Structure and Dynamics}

\begin{itemize}
\item  Light nuclei---ab initio structure calculations with
Faddeev and Green's function Monte Carlo methods;
\item  Halo nuclei---prediction of unusual properties due to neutron
pairing in a low-density environment;
\item Interacting Shell Model---promise for a Monte Carlo
technique to overcome the $N!$ problem;
\item Heavy Nuclei---nuclear shapes and single-particle structure
from Hartree-Fock-Bogo\-liubov theory.
\end{itemize}

\subsubsection{Towards the Quark Structure of Matter}

\begin{itemize}
\item Understanding how to measure
the complete set of quark and gluon distributions in hadrons;
\item Providing evidence for a major role of instantons
in the QCD vacuum and in light-quark hadronic structure;
\item Calculation of properties of mesons and baryons in lattice gauge
theory, including masses, radii, and scattering properties;
\item Applications of chiral perturbation theory to low-energy hadron
dynamics.
\end{itemize}

\subsubsection{The Phases of Nuclear Matter}

\begin{itemize}
\item Validation of one-body transport dynamics for heavy ion collisions
at intermediate energies;
\item Development of phenomenologically successful relativistic hadron
cascade models for heavy ion collisions at the AGS and SPS;
\item Development of partonic cascades predicting the formation of
thermalized high-density matter as required for a quark-gluon plasma
at RHIC and LHC;
\item Improved theoretical understanding of signatures for the QCD
phase transition, including charmonium, jets, lepton pairs, and photons.
\end{itemize}

\subsubsection{Weak Interactions and Nuclear Astrophysics}

\begin{itemize}
\item A more precise understanding of solar neutrino flux uncertainties
and strengthening of the evidence for neutrino mixing;
\item A qualitative understanding of the enhancements seen in
parity-violating compound nucleus reactions;
\item Identification of the likely source of r-process nuclei in the
supernova envelope and the role of neutrino spallation in nucleosynthesis;
\item Stellar models including convection and neutrino physics which
show that collapsing stars can explode in supernovae.
\end{itemize}

\subsubsection*{Acknowledgments}

We wish to express our sincere gratitude to those colleagues who have
helped us in producing this report, especially,
A.B. Balantekin, B.A. Brown, P. Danielewicz, D. Feng,
J. Ginocchio, W. Haxton, F. Iachello, X. Ji, M. Musolf, W. Nazarewicz,
R. Perry, J. Randrup, and B. Serot.
\newpage

\section{Nuclear Structure and Dynamics}

The past five years have seen impressive development of
quantitative theoretical tools to describe and predict nuclear structure
properties.  The information sought depends on the kind of nucleus
studied, and the theoretical techniques are correspondingly
diverse.  In particular, few-body methods, the microscopic
shell model many-body methods, and the phenomenological
application of dynamical symmetries have seen much progress.

\subsection{Microscopic Theory}

\subsubsection{Light Nuclei}

For three-particle nuclei, the Faddeev equations are the method
of choice, and the theoretical precision has reached a point
where the data on reactions in the three-particle system
can be used to refine our knowledge of three-body forces and
of the neutron-neutron interaction.

For nuclei somewhat larger, $A \approx 4-7$, significant advances
were made with Green's function Monte Carlo methods, which allow one
to obtain accurate wave functions and energies from a
trial wave function starting point.
Accurate (called ``exact'' when the Monte Carlo sampling error is
reduced to below levels of interest) solutions were obtained for
the ground states of $^4$He, $^6$He,
$^6$Li, $^6$Be, the $3^+$ and $0^+$ excited states of $^6$Li and the $3/2^-$
and $1/2^-$ scattering states of $^5$He with most realistic models
of two- and three-nucleon interactions. In the near future
it appears possible to calculate several seven and eight nucleon states
with available parallel computer platforms. Calculations of
the $^{16}$O ground state may also be possible with
somewhat less accuracy. Such exact solutions provide detailed
information required to study nuclear forces and nuclear structure.

The quantum Monte Carlo method can also provide information to
analyze electron-nucleus scattering experiments, electroweak
radiative capture and other reactions.  This is obtained through
the Laplace transform of the linear response function.  Pioneering
calculations of the response have been made which resolve the
long standing problem of understanding
the longitudinal and transverse electromagnetic responses
of $^4$He observed at Bates and Saclay laboratories. The nuclear
interactions modify both these response functions, and the pair currents
due to pion exchange interactions enhance the transverse response function
as illustrated in Fig. 1. In the near future it will be possible
to calculate the response functions of nuclei to other probes as well.

\begin{figure}
\def\epsfsize#1#2{0.8#1}
\centerline{\epsfbox{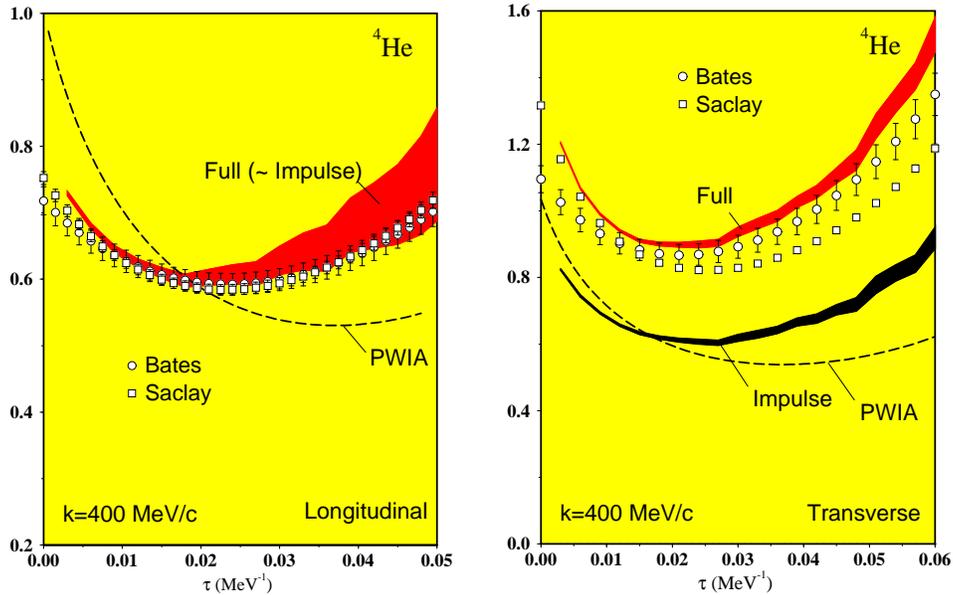}}
\caption{
The ratio of the Laplace transform of the response function of $^4$He to
that of four free nucleons. The agreement of the full calculations
with the data provides a sensitive test of the role of nuclear
interactions and meson exchange currents, especially in the case of
the transverse electromagnetic response function.
}
\end{figure}

\subsubsection{Halo Nuclei}

The recent experimental accessibility of nuclei near the neutron drip
line has engendered much theoretical work to describe the properties
of these nuclei.
In particular, the nucleus $^{11}$Li has emerged as
a paradigm for systems marginally bound by the neutron-neutron
pairing interaction.  Due to the low average density of the
neutron halo, ab initio calculations are simpler.
The weak binding of $^{11}$Li and the lack of binding in the
neighboring nucleus $^{10}$He is explained by realistic
modeling of the interaction between valence neutrons. Many of the
measurable properties around $^{11}$Li were predicted by theory
and confirmed by experiment.  In particular, the presence of
a very soft quasiresonance in the excitation spectrum and the
influence of the neutron pairing correlations
on break-up reactions are quantitatively described by prior theory.
The conclusion one can draw
from this interplay of theory and experiment is that our current
understanding of the nuclear force is adequate for the theory
of neutron pairing in a low-density environment.

\subsubsection{Shell Model}

Shell model calculations for light nuclei can now be made using all
many-particle states within a large oscillator space.  This
so-called ``no-core'' model space simplifies the determination of
an effective interaction from an underlying realistic force
via the Brueckner G-matrix theory.  The method is thus ab initio
beginning from a realistic interaction with a well-defined
approximation scheme.

For many years the systematic application of the shell model
basis to open-shell nuclei has been stalled at nuclei
lighter than the $A=40$ magic number, due to the factorial
growth of configurations in open-shell systems.  A new technique
shows promise to overcome this problem.  The strategy
is to sample the relevant configurations rather than try to include
them all explicitly.  The method is most appropriate to study properties
of thermal ensembles, and it relies on an extrapolation of the nuclear
interaction to allow effective Monte Carlo sampling. An application to
the Gamow-Teller strength of nuclei in the $Z=26$ region demonstrates
that the method can produce results of interest for nuclear
physics input to supernova dynamics and element production.

The possibility to go to much larger shell model bases underlines
the problem of determining the effective interaction appropriate
to a given basis.  For nuclei near closed shells, progress has
been made on finding a consistent global effective interaction
with a range extending from mass number $A=40$ to $A=208$.
However, much work remains to be done in the search for a
realistic global interaction.

\subsubsection{Mean Field Theory}

Mean field theory provides the starting point for a microscopic theory
of heavy nuclei.  With the advent of fast computers, self-consistent mean
field calculations can be performed eliminating all artificially
imposed symmetries.  These Hartree-Fock-Bogoliubov (HFB) models
are remarkably accurate in describing the global and single-particle
properties of medium-mass and heavy nuclei.  Several versions of the
input Hamiltonian have been developed, emphasizing simplicity, exact
treatment of the exchange interaction, or relativity.

In one recent example, the quadrupole moment of a high spin isomer,
predicted in HFB, was measured for the first time.
The nucleus $^{178}$Hf was found to have a
quadrupole moment of 7.2 eb comparing very well with the
theoretical value of 7.3 eb.  Other, more exotic
shapes are also predicted by mean field theory. For example,
the octupole deformations in barium, radium and thorium isotopes is
rather well described by HFB.  Mean field theory
gives the best guidance on the energies of heavy nuclei at the
farthest limits of stability.  The recent observation of a
nucleus with charge $Z=110$ confirmed predicted binding energies
to an accuracy in the range $0.1-0.3$ MeV.  Finally,
the puzzle of identical bands has been addressed within the
framework of relativistic mean field theory.

Calculations of ground state properties in stable and exotic
nuclei, and at high spin states, indicate the importance of density
dependent pairing forces. Significant progress has been made in the
description of collective dynamics. Examples are many-dimensional
self-consistent calculations based on the generator coordinate method.

\subsection{Collective Models}

\subsubsection{Interacting Boson Model}

The Interacting Boson Model (IBM) has been further extended to cover
odd-odd nuclei and broken pair states. Detailed
predictions are now available for all types of nuclei (even-even, even-odd
and, to some extent, odd-odd) extending to nuclear species far from
stability. These predictions, which include beta-decay strengths and
binding energies, can be tested at radioactive beam facilities and can
form the input for calculations of astrophysical interest.

Calculations of the spectra of medium mass and heavy nuclei with realistic
interactions have been completed by making use of the bosonization method.
This method has many similarities with techniques used
in condensed matter physics and in its present version includes $S$-, $D$-
and $G$-pairing.  The results of the calculations agree with experiment
within few percent.  Recent progress in fermion to boson mapping may
lead to progress for very deformed systems as well.

The robust and elegant concept of dynamical symmetry is one of the
central threads of physics. One of its original manifestations has been
in very deformed nuclei with axial symmetry. For instance, the O(6)
symmetry, which occurs in both the interacting boson model and the
fermion dynamical symmetry model, is unique to nuclear physics
and has not been observed in other systems with collective motion,
such as molecules. Recently many bands of the U(5) dynamical symmetry
has been observed. Furthermore, dynamical supersymmetry, first
observed in odd-even nuclei, and not observed in any other physical
system to date, may help bringing an order to the complex spectroscopy of
odd-odd nuclei. Also, dynamical symmetry gives an elegant
classification of neutron-proton symmetric and antisymmetric states.

Relatively strong magnetic transitions have been observed at about 3 MeV
excitation in heavy nuclei, and the strength is found to be proportional
to the B(E2) strength in the nucleus.  This systematics has
been explained from several points of view.  It is implicit
in the original collective model of the excitation as a
``scissors'' mode oscillation between deformed proton and
neutron ellipsoids.  Recently, sum rules have been derived
to explain this proportionality.  In a sum rule derived
from the Interacting Boson Model, the dipole strength is
proportional to the number of quadrupole bosons in the
ground state.  In a schematic shell model with
quadrupole-quadrupole interactions, a energy-weighted sum
rule has been derived that also exhibits the proportionality
between magnetic and quadrupole transition strength.

By combining the reaction mechanism and the nuclear structure in a
unified manner within the IBM, higher order coupling terms beyond
the Born approximation are calculated and lead to improved
agreement between barrier distributions calculated in the IBM and
measured in fusion reactions.
In particular it has been shown that the barrier distribution has a
marked dependence on the nuclear shape, a prediction which will
inspire measurements on transitional nuclei.

\subsubsection{Superdeformation}

Since the original discovery of the first superdeformed band,
identical bands have been seen in the neighboring superdeformed nuclei
having an additional nucleon. A partial explanation for this
striking fact stems from a discovery made many years ago.
In heavy nuclei certain doublets of single particle orbits are almost
degenerate. Although these orbits are not spin-orbit doublets, they
have single-particle angular momentum $J$ differing by 1/2, which is
called the pseudo-spin. If this pseudo-spin does not participate in the
collective rotation, the rotational energy is the same in the
neighboring nuclei provided the moment of inertia does not change.
The origin of this symmetry is just beginning to be understood.
Why the moments of inertia are identical to better
than the one percent change expected from the variation in mass and
radius still remains to be clarified.

Recent experiments have detected a $\Delta J = 2$ staggering in the
spectrum of a superdeformed band, and it has been
suggested that this may be evidence for a change from an axially
symmetric shape to a non-axial shape with $C_{4v}$ symmetry
(invariant under rotation of 90$^o$ about the original symmetry axes).
Initial calculations with an intrinsic Hamiltonian
show that this staggering can be reproduced with a deformed shape
which includes a small hexadecapole deformation of the form $Y_{44}
+ Y_{4,-4}$, although precise calculations are not possible at this
time.

\subsubsection{Nuclei Far from Stability}

In order to have states with good isospin for nuclei with neutrons
and protons filling the same major shell, a collective neutron - proton
pair must be included along with the neutron pair and proton pair to
complete the isospin triplet. This IBM-3 model, previously
applied only to light nuclei, has been applied to the nuclei with
$N \approx Z \approx 40$; a comparison of the calculated and measured
spectrum of $^{80}$Sr shows good agreement.
Although the nuclei in this region are very deformed, the
deformation is not axially symmetric but gamma unstable, and with
good O(6) symmetry. This type of collective motion is unique to nuclear
physics and needs to be studied more. Present studies suggest that
nuclei with $N \approx Z$ and the valence shell nearly half filled will
have O(6) symmetry. Also for $N \approx Z$ magic numbers may change due to
large neutron-proton interaction, as we have seen for $^{80}$Zr.
An interesting question is whether $^{100}$Sn will be doubly magic.

\subsection{Challenges and Opportunities}

One of the challenges which emerges from these past developments is
to exploit the full power of emerging computer technology to
solve important nuclear many-body problems which are now accessible
for the first time. Unrestricted
HFB calculations of nuclei far from stability
will play an essential role in understanding the physics to be
explored with radioactive ion beams. Green's function Monte
Carlo solutions could be extended beyond $A=6$ nuclei, and offer
the potential for a microscopic understanding of both
ground state properties and response functions of light nuclei.
Given the progress in proving feasibility of Monte Carlo shell model
calculations using simple phenomenological interactions,
another major challenge will be
to develop microscopic effective shell model interactions whose
physics content is commensurate with the large numerical effort
required for these calculations.
\newpage

\section{Towards the Quark Structure of Nuclei}

Given the major investment which has been made in CEBAF, BATES, and
RHIC as well as in nuclear physics experiments at HERA, SLAC, and
Fermilab, understanding the role of quark and gluon degrees of freedom in
hadrons and nuclei has become a major focus of contemporary
theoretical nuclear physics.

\subsection{Quark and Gluon Distributions in Hadrons}

Deep inelastic lepton scattering has provided extremely important
but as yet incomplete information about the quark/gluon structure
of the nucleon.  For example, the measurement of the spin-independent
quark distribution $f_{1}(x)$ shows that only half of the energy plus
momentum of the nucleon is carried by quarks and the longitudinal
spin-dependent distribution $g_{1}(x)$ shows that quark helicity carries
very little of the nucleon spin and suggests that the strange quark
sea may be strongly polarized.  Recently, a careful, systematic
analysis has provided a complete classification of all the quark
and gluon distribution functions which can be defined in the nucleon
with precise, quantitative relations of these distributions to the
structure functions which can be measured experimentally in high energy
scattering.

Based on this understanding, theorists have proposed new experiments
to measure previously unknown quark and gluon distributions and several
of these proposals
are already being carried out or planned.  The theoretical idea
of using pion production by a polarized electron beam on a transversely
polarized proton target to measure the chirally odd transverse spin
distribution $h_{1}(x)$ is being carried out by the Hermes collaboration
at HERA and will be the first measurement of this new spin dependent
quark distribution.  The Drell-Yan process in polarized $\vec{p} \vec{p}$
collisions at RHIC has been proposed as another way to measure $h_{1}(x)$
and is an important part
of the experimental program proposed by the RHIC Spin Collaboration.
Direct photon production in proton-proton collisions is
a sensitive probe of the gluon distribution, since the scattering of a
gluon from a quark to produce a photon plus a quark involves the gluon
distribution at tree level.  Thus, the theoretical observation that
direct photon production from longitudinally polarized protons would
measure the helicity polarization of gluons $\Delta g$ has led the
RHIC Spin Collaboration to propose measurement of $\Delta g$ as well.
These are important examples of the intellectual
leadership QCD theorists can provide in exploiting the full potential
of the emerging facilities in our field.

\subsection{Structure of the QCD Vacuum and Hadrons}

Significant evidence has been found that instanton-induced forces
play a dominant role in light quark physics.  At the phenomenological
level, an instanton liquid model economically accounts for the gross
behavior of light quark systems.  It accounts in detail for the behavior
of point-to-point correlation functions of hadronic currents in the
QCD vacuum, it generates the quark condensate and pions, it reproduces
the gross properties of hadrons and accounts for the major features
of glueballs.

These phenomenological features have been confirmed by numerical
solution of QCD on the lattice.  Analysis of the instanton
content of lattice calculations confirms the phenomenological average
instanton size of $\frac{1}{3}$ fm and density of about 1 fm$^{-4}$.
Perhaps the strongest evidence for the dominant role of instantons
arises from comparison of lattice calculations which correctly include
all the excitations of the gluon degrees of freedom with those in which
essentially all excitations except for instantons have been removed as
shown in Fig. 2.  The result is that the gross distribution of quarks in
hadron ground states and the behavior of point-to-point hadron vacuum
correlation functions are hardly affected by removing all other degrees
of freedom.  Thus, in considering light hadron physics, the task of
understanding nonperturbative QCD is greatly simplified by identifying the
instanton configurations which saturate the QCD vacuum, which allows one
to ignore summing over all the other gluon degrees of freedom which in the end
do not play a major role.

\begin{figure}
\def\epsfsize#1#2{0.65#1}
\centerline{\epsfbox{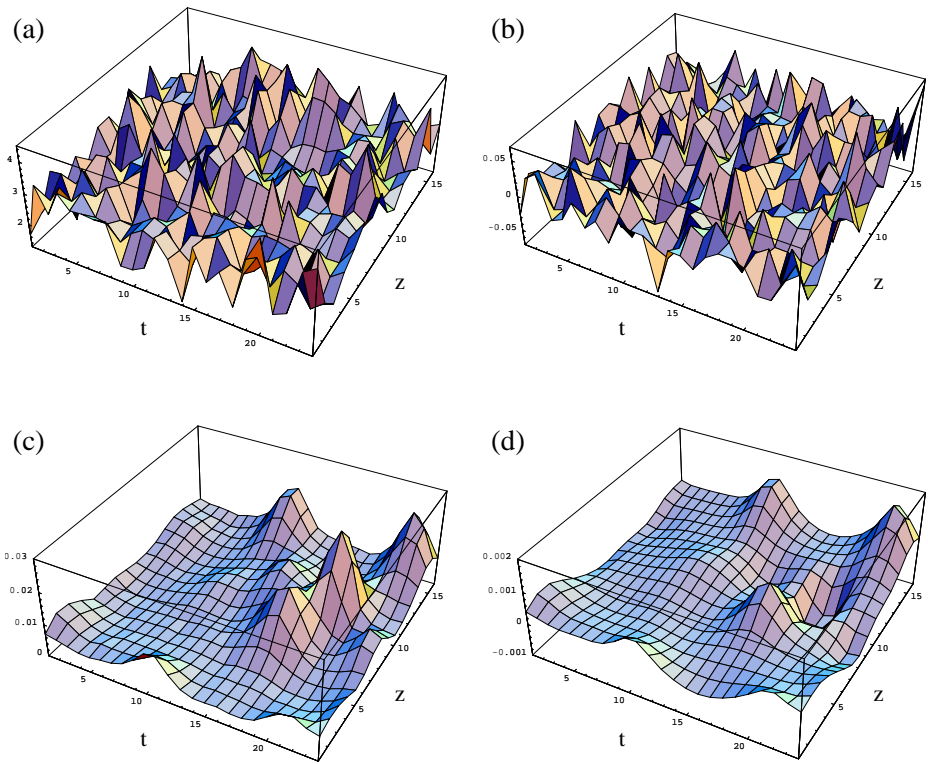}}
\caption{
The instanton content for a typical slice of a gluon
configuration calculated in lattice QCD at fixed $x$ and $y$ as
a function of $z$ and $t$. The left column shows the
action density $S(1,1,z,t)$ including all gluon contributions (a)
and after removing all contributions except for instantons (c).
The right column shows the topological charge density
$Q(1,1,z,t)$ before (b) and after (d) removal of non-topological
excitations, revealing the presence of 3 instantons and
2 anti-instantons.
}
\end{figure}

\subsection{Lattice Gauge Theory}

Numerical solution of QCD on a discrete space-time lattice is the only
known way to solve, rather than model, QCD. Starting with a Euclidean path
integral, the Gaussian integral over quark fields is performed analytically,
yielding  an integral over gluon fields which is evaluated numerically using
Monte Carlo sampling. The so-called quenched or valence approximation, in
which the numerical calculation is greatly simplified by omitting the
non-local fermion determinant, corresponds to omitting all quark-antiquark
excitations out of the Fermi sea.

One major accomplishment over the past few years has been the successful
calculation of a broad range of hadron observables in the quenched
approximation. These include masses, form factors,
magnetic moments, moments of structure functions, meson decay constants,
the quark condensate $\langle\bar\psi \psi\rangle$,
vacuum correlation functions of hadron currents,
pion-pion and pion-nucleon scattering lengths, and the
pion-nucleon sigma-term $\sigma_{\pi N}$.  In the best
cases, such as masses, agreement with experiment is at the 5\% level,
and systematic effects due to lattice spacing, lattice volume, and finite
quark mass are well understood and controlled. A second major development
has been the demonstration that in cases where one knows that the Fermi
sea should contribute significantly, such as decay constants
and the sigma-term, inclusion of the fermion determinant yields large
corrections which significantly improve agreement with experiment.

Lattice QCD
has also provided important insight into the structure of hadrons
and the QCD vacuum. In addition to elucidating the role of instantons
in quark propagation, numerical calculations also yield direct
evidence for the role of monopoles in confinement and provide a
quantitative measure of the contribution of gluons to hadron wave
functions.

\subsection{Light-Front QCD and Dyson-Schwinger Equations}

Lattice calculations are still limited to ground or low-lying excited
states and equilibrium thermodynamic quantities at zero chemical
potential, which excludes much of nuclear physics. The development of
other  analytic techniques  and numerical approaches that provide
systematic approximations is therefore of great importance.

One interesting approach is light-front QCD, which
offers the advantages of a simple vacuum and close correspondence to the
quantities
measured in deep inelastic scattering.
Present computational schemes employ cutoffs that
violate explicit rotational invariance and gauge invariance, forcing
symmetry-breaking terms to appear in the Hamiltonian.
Using renormalization group techniques, progress has been made on
these renormalization problems and in seeing how
asymptotic freedom and confinement emerge in this theory.
Since light-front field theory effectively reduces the number of non-trivial
dimensions by one, it has proven to be a powerful numerical tool
in $1+1$-dimensional gauge theories.
The light-front framework is also useful phenomenologically,
providing a boost invariant
tool for generalizing nonrelativistic quark model
calculations to intermediate energies.
There has been substantial
progress in applying light-front constituent quark
models to the study of electromagnetic form
factors and reactions.
Dynamic restoration of rotational and gauge invariance broken in these
models remains an outstanding challenge.

Another promising computational approach to QCD is provided by the
Dyson-Schwinger equations, which can be truncated without breaking gauge
invariance or chiral symmetry. It has been shown that the
three-gluon vertex is sufficient to drive confinement, while the known
perturbative ultraviolet behavior of QCD is explicitly
recovered. It is presently necessary to parameterize the
low-momentum behavior of the gluon propagator since it has not been
calculated analytically or numerically on the lattice, but reasonable
parameterizations reproduce $\pi$-$\pi$ scattering
and the electromagnetic form factor
of the pion.

\subsection{Effective Low-Energy Theories of QCD}

Effective Chiral Lagrangians are the result of transforming QCD from its
underlying quark and gluon degrees of freedom to effective
hadron degrees of freedom.
The broken chiral symmetry manifested by these building
blocks acts to constrain the strong interactions, and the resulting chiral
perturbation theory provides a quantitative
calculational framework. A wide
variety of observables such as the electric and magnetic polarizabilities of
the neutron and proton, the $\gamma p \rightarrow \pi^0 p$
amplitudes near threshold,
and many properties of pions have been calculated,
yielding agreement with experiment in the pion sector.
Recent measurements of the proton polarizabilities at
Saskatoon, Illinois, and Mainz,
and the neutron electric polarizability at ORNL
are in agreement with these calculations.  The novel predictions for
near-threshold neutral-pion photoproduction
from the proton are currently being tested experimentally.

An interesting new development is the application of techniques from
the heavy quark effective theory to include baryons in a systematic
manner in the chiral perturbation expansion.  This
approach has been applied to predict electromagnetic properties of the
ground state decuplet baryons, such as magnetic moments and
electromagnetic decay widths, which will soon be tested
at CEBAF.

Recently, chiral perturbation theory has also been applied to nuclear
forces and was shown to provide a quantitative scheme for
understanding the fact that three-body forces are significantly weaker
than two-body forces and that four-body forces are almost negligible.
It also accounts for the fact that charge independence breaking in the
nuclear force is larger than charge-symmetry breaking and
for the suppression of heavy-meson exchange currents in weak and
electromagnetic interactions.

Several $np$ charge symmetry experiments were stimulated in part by
the efforts of theorists to delineate possible sources
of charge-symmetry-breaking and charge-independence-breaking
in the NN interaction.  The reexamination of $\rho-\omega$ mixing,
particularly its evolution with the momentum carried by the
meson, has been the focus of recent work.

\subsection{Relativistic Nuclear Many-Body Theories}

Relativistic hadronic field theories can be constrained to fit low-energy
phenomena observed in ordinary nuclei and then extrapolated to more extreme
situations of density, temperature, flow velocity, and four-momentum
transfer, such as those that will be produced in experiments at
RHIC and CEBAF. Relativistic approaches provide a natural way to
incorporate the symmetries of QCD, such as chiral symmetry and broken scale
invariance, at the hadronic level.

The relativistic quasipotential approach to nuclear matter, known as
Dirac-Brueckner-Hartree-Fock theory, incorporates large scalar
and vector self-energies into the nucleon wave functions.
The effective scalar field represents the correlated two-pion exchange
between nucleons in a nonlinear realization of the chiral symmetry.
The introduction of a quartic self-interaction of the vector field
has been found to be important for accurately modeling the
self-energies and consequent nuclear properties.
This approach introduces a density dependence into the NN interaction that
goes beyond what is included in nonrelativistic Brueckner theory and makes
it possible to simultaneously fit both the NN phase shifts and
the nuclear matter equilibrium point at first order in the low-density
expansion.

To extend these results to finite nuclei,
one parameterizes the density dependence of the self-energies and
then introduces this density dependence into the hadronic Lagrangian by
allowing the meson-baryon couplings to be functionals of the baryon fields.
The resulting Dirac-Hartree equations naturally include rearrangement terms.
The results for rms radii, binding energies, and charge distributions of
closed-shell nuclei are quite successful.

A recent conceptual insight is that the phenomenologically important
term associated with so-called ``Z-graphs'',
which mix the free positive- and negative-energy Dirac wave functions,
is a natural consequence of Lorentz covariance.
This term is therefore model independent and
should be present whether the nucleon is composite or not.
At present, nothing is known about the origin
of the important spin-orbit terms, which are also signatures
of the Dirac approach, but work is underway to see if these
can also be generated in a model-independent fashion.

\subsection{Challenges and Opportunities}

The most important challenges in this field are to obtain a fundamental
description of hadron structure through the quantitative solution of QCD
and to improve our understanding of the quark structure of nuclei through
both explicit QCD solutions and models that are consistent with low-energy
QCD. Recent developments provide unprecedented opportunities for progress.

To date, the primary focus in studying quark and gluon distributions in
hadrons experimentally
has been the so-called leading twist distributions which dominate
structure functions at the highest momentum transfer. With the advent of
CEBAF, the opportunity now exists for precision exploration of the
sub-dominant higher twist distributions which become measurable at low
$Q^2$. These distributions display the effects of
coherent parton scattering beyond the Feynman parton model, and
directly measure the QCD initial and final state interactions at the quark
and gluon level. Theoretical challenges include subtle ambiguities in
separating distributions of different twist because of infrared physics and
learning to calculate and understand the role of important quark
and gluon configurations revealed by higher twist matrix elements.

With the emergence of teraflops-scale computers, improvements in
algorithms, and the discovery of more accurate discrete approximations
to the continuum action, definitive lattice calculations of
many important observables will be possible for the first time.
Moments of spin-dependent structure functions and transition form factors
can be calculated accurately in the next few years and will be
directly relevant to measurements in the same time frame at HERA and
CEBAF.  Definitive calculation from first principles of the low
energy parameters of chiral perturbation theory will elevate what is
now a limited phenomenology with a large number of free parameters to
a far more fundamental and quantitative theory. The quest for insight
into the role of topological structures such as instantons will also
advance significantly with the possibility of studying instantons in full
unquenched QCD with lattices which provide an accurate approximation
to the continuum limit. A major challenge to the field is to devote the
computational resources and effort required for the solution of QCD and
understanding the data from new experimental facilities.

Light-front QCD poses substantial theoretical challenges
associated with lack of rotation
invariance, renormalization, and understanding the relation between
observables in the lab and infinite momentum frames.
A crucial question is whether
the advantages outweigh the renormalization
difficulties in $3+1$-dimensional QCD. The method of truncated
Dyson-Schwinger equations is now poised for application to a large
variety of problems in hadronic structure and dynamics. A major
challenge for this approach will be constructing a controlled,
systematic framework for using solutions from lattice
gauge theory to characterize
the low-momentum gluon propagator
and effective vertices.

It will be important to determine the extent to which effective chiral
Lagrangians
can provide a tractable framework for calculating
the nuclear force, with all of its richness and complexity.
Quantitative progress would further solidify calculations in few-nucleon
systems
and light nuclei.
The development of a reliable, quantitative low-energy effective theory
based on hadrons will be important for understanding CEBAF experiments and
will depend on the identification of relevant expansion parameters for
strongly interacting, relativistic many-body problems.
Such a theory would be extremely valuable in understanding the novel forms
of dense matter which are of astrophysical interest, such as the
possibility of pion or kaon condensation occurring in neutron stars.
It would also permit calculations of hadronic properties as functions of
density and temperature that can be used to extrapolate to extreme
conditions.

Even with expected significant progress in analytical and numerical solution
of QCD, there are a number of expected developments in low-energy
hadron spectroscopy at CEBAF and other laboratories for which
quantitative solutions will be impractical and
hadron models will therefore play a crucial role. Enlightened model
building must address at least two classes of questions. One is why
states predicted by otherwise successful models are not observed
experimentally. The other is understanding the precise nature of
hadron resonances, for example the structure of the Roper resonance, N(1440),
the mixing of the S$_{11}$ resonances, N(1535) and N(1650), and whether
states like the scalar mesons f$_0$(975) and a$_0$(980) can be understood
as quasi-molecular states. Quark models, flux tube models, and algebraic
models offer the potential of complementary insights.
\newpage

\section{The Phases of Nuclear Matter}

An important goal of nuclear physics is to observe and measure
the possible phases that nuclear matter can assume under
various conditions of energy and density.  Heavy ion collisions
are the experimental means to produce such phases, but the
measurements can be meaningfully interpreted only if a
reliable transport theory is available to relate the observables
back to the fundamental dynamics.  There has been substantial
progress in the last five years in the development of transport
theory and applications to heavy ion collisions.  The specific
theoretical models emphasize different aspects of the dynamics,
depending on the energy regime studied.  At low energies,
one-body transport with mean field and collisional dynamics
has proven useful.  At higher energies, the transport models
necessarily probe the quark and gluon degrees of freedom within
the colliding nuclei.

\subsection{Low energies: One-body transport and beyond}

Transport theory at nonrelativistic energies is based on
a one-body approximation in which the nucleon density in
phase space is evolved as a classical variable.
Quantum physics enters implicitly through the
parameters of the interaction and explicitly through
Pauli-blocking of nucleon-nucleon collisions.
This approach has been successful in describing a number of experimental
observables. Angular distributions for nucleon production are reproduced
at the level of 10 to 20 percent for beam energies from tens of MeV to
several GeV. The production of light composite nuclei, such as deuterons,
and of pions is also well described. At energies above a GeV,
nucleon resonances are important as intermediaries for the
production of $\eta$ mesons, kaons and antiprotons.
Proton-proton correlations, which provide
information on the space-time evolution of the
colliding nuclei, also agree reasonably with observation.

An important goal of heavy ion reaction studies is to
measure the nuclear equation of state. The collective sideways flow in
an off-center collision is an important observable since it
is sensitive to the equation of state at higher densities, as well as
to the effective cross section between nucleons in the nuclear medium and
to the momentum dependence of the mean field. From
flow data on medium-heavy nuclei at a bombarding energy of
400 MeV/n, one can constrain the nuclear incompressibility, with
current theory yielding $ \kappa < 250 $ MeV.

An intriguing prediction of transport theory is that a
doughnut-shaped nucleus can be produced in nuclear collisions
under certain circumstances.  The collision has to be nearly
head-on to preserve the symmetry of the shape, and the
energy has to be in a rather small window around 60 MeV/n to
produce the large-scale deformation without blowing the
combined system apart.  The result of a transport model calculation is shown
in Fig. 3. An outstanding challenge is to confirm this precidtion
with a clear experimental signature.

Another goal of studies in the medium energy range is to
find  observable effects of the gas-liquid phase transition.
To understand multifragmentation yields, the average Boltzmann collision
term is augmented by a fluctuating Langevin term
and the resulting theory describes an ensemble of
one-body densities.  Simplifying analytical
approximations have recently reduced the numerical
effort to the point that the method
is now practical for realistic calculations.

A more quantum mechanical approach to treat the regime of large
fluctuations is to constrain the A-body wave function to be a
Slater determinant of Gaussian single-particle wave packets,
yielding a set of dynamic equations for the positions of
the wave packets and their widths.  This approach has
given a new understanding of the reactions involving
relatively light ions, but remains impractical for heavy systems.

\begin{figure}
\def\epsfsize#1#2{0.8#1}
\centerline{\epsfbox{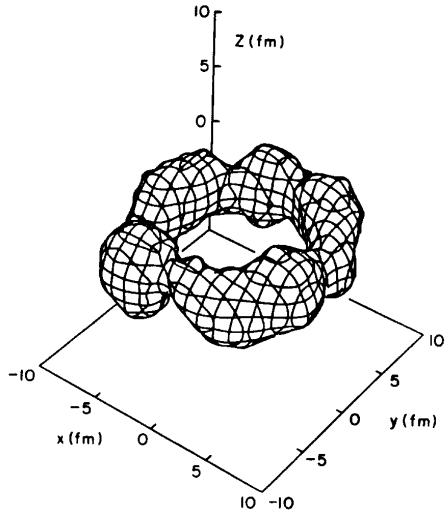}}
\caption{
A doughnut-shaped nucleus predicted by transport theory.  This
shape was produced simulating a collision of $^{93}$Nb on $^{93}$Nb
at a beam energy of 60 MeV/n.  The contour shows the surface
at 1/3 nuclear matter density at a time $t=160$ fm/$c$ from
the start of the collision.}
\end{figure}

\subsection{Hot and Dense Hadronic Matter}

Experiments with relativistic heavy ions at the Brookhaven AGS and
the CERN-SPS aim to investigate the properties of dense and highly
excited hadronic matter, and possibly to study the onset of chiral
symmetry restoration in baryon-rich dense matter. Substantial progress
has been made since the last Long Range Plan on the theoretical
description of nuclear collisions at AGS and SPS energies, and on
the problem of medium effects on hadrons in hot and dense hadronic
matter.

A number of relativistic transport models has been developed during the
past six years which model collisions of nuclei at relativistic energies
as cascades of colliding hadrons and take into account rescattering.
These are implemented in codes such as VENUS, RQMD, and ARC.
At AGS energies, these codes are based on almost identical
assumptions, i.e. hadrons interacting pairwise with free-space cross
sections determined by experimental data from $pp$ collisions and quark
model extrapolations, with the possible addition of a density dependent
mean field. At the higher CERN energies, the codes make use of $pp$ collision
phenomenology, mostly within the framework of the Lund string model.

Overall, these models have enjoyed impressive success in describing and
even predicting a wealth of inclusive particle spectra measured at
Brookhaven and CERN with few, if any, adjustable parameters. This has
allowed theorists to obtain a detailed picture of the reaction dynamics,
especially at AGS energies, in terms of a dense gas of meson and baryon
resonances reaching baryon densities up to $9\rho_0$, where $\rho_0$ is
the density in the nuclear ground state.  As an example,
Fig. 4 shows the evolution of the baryon density in Au+Au, Si+Au,
and Si+Si collisions at AGS energies.  The rising part of the curves
describes the rapid compression and heating of nuclear matter upon impact,
followed by a slower, almost isentropic expansion and ending in a freeze-out
of hadrons after about 15 fm/$c$. A large fraction of the nucleons is
temporarily in excited states which form a major source of the pions
created in these collisions. The matter appears locally equilibrated in
the later stages of the reaction, when densities are still far above
normal nuclear density, providing strong arguments for the validity of
thermal models for the spectra of emitted particles.

\begin{figure}
\def\epsfsize#1#2{0.45#1}
\centerline{\epsfbox{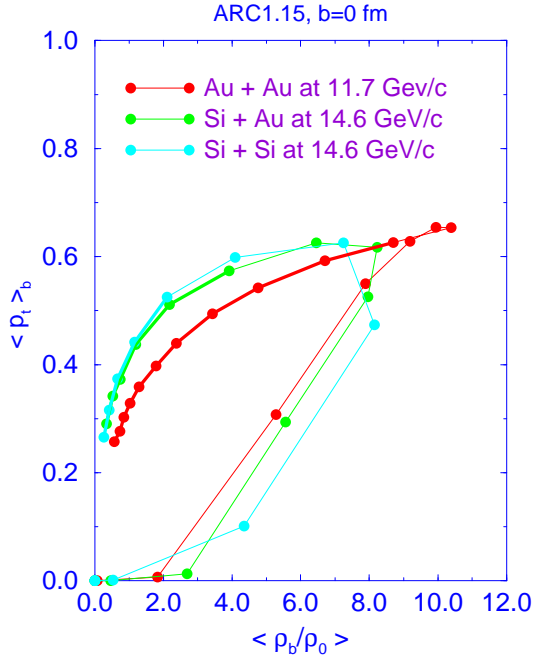}}
\caption{
Time evolution of the baryon density and average transverse momentum of
hadrons in the center of central Si+Si, Si+Au, and Au+Au collisions at
the AGS as predicted by the ARC code. All curves begin at the origin
and are to be followed counter-clockwise. The dots indicate equal time
intervals of length 1 fm/$c$. The baryon density is calculated in
the local center-of-mass frame. The highest densities exceed $9\rho_0$.
}
\end{figure}

Theoretical expectations are that the properties of hadrons, such as mass
and decay width, should be modified in the presence of a dense and highly
excited environment, even before the phase transition to a deconfined
quark-gluon plasma. At temperatures much less than $T_c$ it is most
efficient to think of the matter as a relatively dilute gas of hadrons.
This may be described by a virial expansion, using low energy
effective Lagrangians, chiral perturbation theory,
or QCD sum rules. Although it is generally agreed
that there is a tendency toward restoration of chiral symmetry at high
temperature or density, that is reflected in an approach to degeneracy
between parity doublets such as the $\rho$ and $a_1$ mesons, predictions
derived from different methods do not coincide. In particular, lattice
results currently indicate little or no change in the screening masses
with temperature except in the immediate vicinity of the chiral phase
transition.

The modification of hadronic properties in the interior of nuclei is
being actively studied using a variety of techniques.  There is strong
evidence that the chiral (quark) condensate inside nuclei is
significantly reduced (30--40\%) from its vacuum value.
Through general scaling arguments, QCD sum-rule methods, and
quark-hybrid models, this partial restoration of chiral symmetry
has been linked to density-dependent phenomena such as mass shifts of
vector mesons and large scalar and vector nucleon self-energies
suggested by relativistic phenomenology.  Density dependent hadron
masses have already been used in transport models for relativistic
nuclear collisions.  The challenge will be to
solidify these predictions and to reconcile different approaches.

\subsection{Ultrarelativistic Heavy Ion Collisions}

\begin{figure}
\def\epsfsize#1#2{0.8#1}
\centerline{\epsfbox{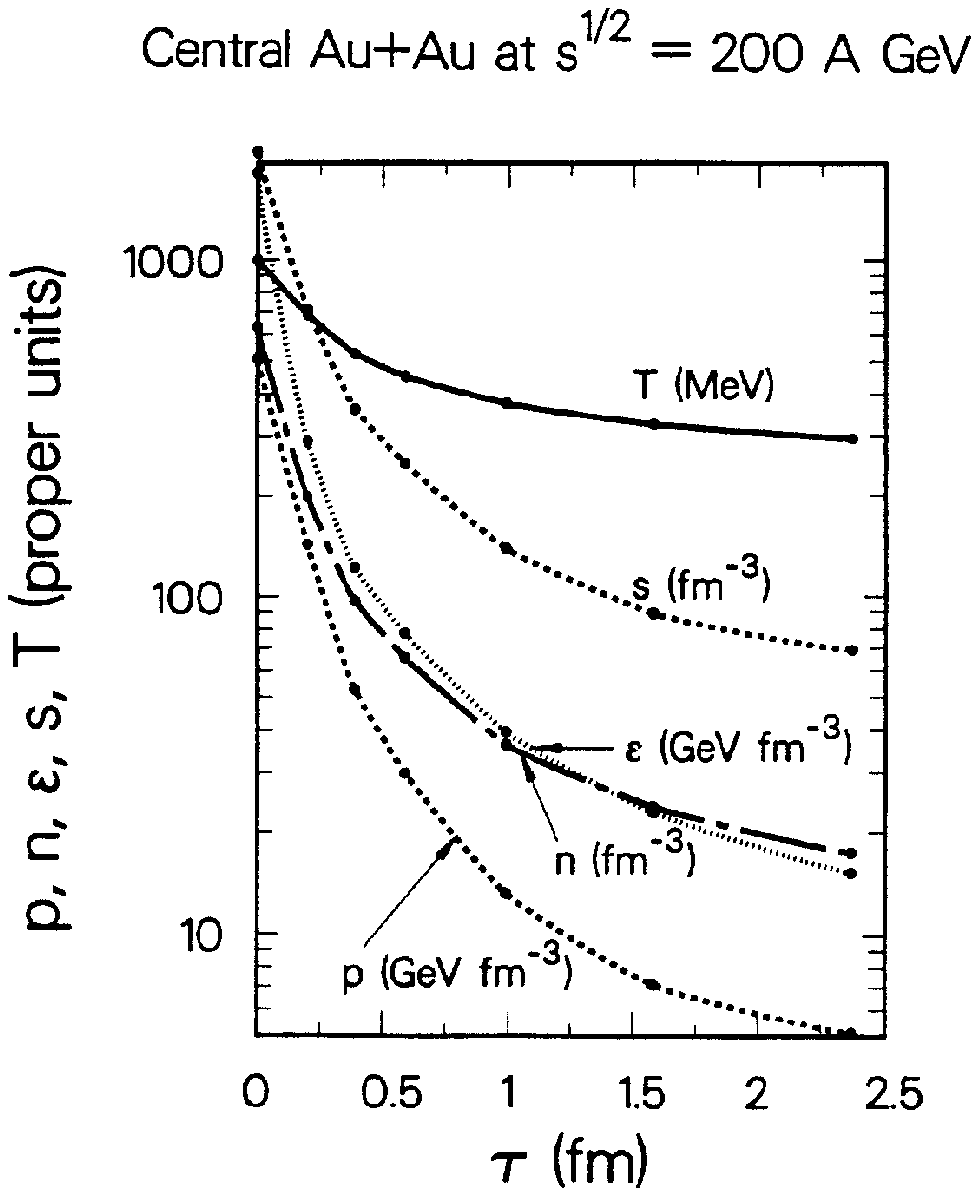}}
\caption{
Time evolution of the energy density $\varepsilon$, local temperature $T$,
and other thermodynamic variables in the central rapidity region during
a Au+Au collision at RHIC energy, as predicted by the parton cascade model.
}
\end{figure}

The past five years have seen significant progress in applications of
perturbative QCD to the problem of thermalization and quark-gluon
plasma formation in relativistic heavy ion collisions at RHIC energies
and beyond. The techniques of renormalization group improved perturbative
QCD, which were developed in the context of jet formation in e$^+$e$^-$
and p\=p collisions, have been applied to describe the earliest
phase of nuclear collisions at RHIC in terms of a partonic cascade.
These calculations have led to a revision of theoretical
assumptions about the initial parameters for the thermalized plasma to be
formed at RHIC toward higher initial temperatures ($T \geq 500$ MeV) and
a high density of gluons (``hot glue scenario''). This,
in turn, has significantly influenced the conceptual design of the major
RHIC detectors, with a stronger emphasis on the measurement of
electromagnetic probes and total charm yield.

Quantitative predictions
for the full space-time evolution of dense QCD matter at RHIC and LHC
energies up to the final hadron distributions are now available,
as shown in Fig. 5. Recently, it has been realized that parton densities
in large nuclei at small Feynman-$x$ may be calculable in the framework
of perturbative QCD. An intense effort is underway to
combine this approach to parton shadowing with medium corrected parton
cascade calculations to provide model-independent QCD prediction for
heavy ion collisions at RHIC.

\subsection{Quark-Gluon Plasma}

Historically, lattice QCD has played a central role in our exploration
of the deconfinement phase transition at zero chemical potential,
which is relevant to the central rapidity regime of relativistic
heavy ion collisions. There is now solid evidence that the transition
temperature with two flavors of light quarks is  $T_c \approx 150$ MeV
and an important new integral technique has been developed to
measure energy and entropy densities. Although the order of the
transition is not yet clear for the physical case of up, down and strange
quarks, in the domain of second-order phase transitions, scaling
analysis of lattice measurements strongly suggests that
the critical exponents are consistent with O(4) symmetry.

Important progress has also been made in the perturbative description of
the equation of state and the transport properties of the quark-gluon
plasma. A resummed perturbation theory for the high-temperature phase of
QCD has been developed, which permits systematical inclusion of screening
effects into an effective action for thermal QCD at momentum scales of
order $gT$ and $T$. This has facilitated quantitative predictions for
photon and dilepton emission from a quark-gluon plasma,
as well as the determination of new transport coefficients such as the
quark and gluon damping rates, equilibration times,
and color conductivity. The resulting picture of a tightly
coupled plasma of quarks and gluons with strongly damped collective
excitations, which is also supported by new results about the chaotic
behavior of the classical nonabelian gauge fields,
confirms the prediction of very rapid thermalization from the parton
cascade models.

In preparation for the experimental program at RHIC, the various
proposed signatures of a quark-gluon plasma have been under intense
theoretical scrutiny.  The energy loss of jets in a quark-gluon plasma
has been calculated and quantitative predictions for jets
propagating through dense matter are available.  The
dissolution of $J/\psi$ and other heavy vector mesons in the
deconfined phase has been extensively studied, and
a comprehensive description of $J/\psi$ and $\psi$ suppression by
hadronic comovers was developed.  A new signature for the
chiral phase transition, the formation of disoriented chiral
condensates (DCC), was proposed and extensively
investigated.  DCC's would produce
highly specific experimental signatures, such as unusual pion charge
ratios, or charge correlations, which should
be detectable by the RHIC detectors.  Hadronic mechanisms for
strangeness equilibration have been studied in the context of cascade
codes, and shown to fail in generating the level of equilibration
already observed at SPS energies, confirming
strangeness enhancement as a promising signature of quark-gluon plasma
formation. Significant progress has also been made in our understanding
of suitable signatures for a long-lived mixed phase, such as lepton
pairs, vector mesons, and two-particle correlations.

\subsection{Challenges and Opportunities}

Fundamental challenges in this field are to develop a quantitative
description of the phase transition from hadronic to quark-gluon
matter, to determine the structure of nuclear matter on both sides of
the transition, and to identify reliable experimental signatures for
chiral symmetry restoration and deconfinement.  An important step
toward that goal is the refinement of quantum transport theory for
nonequilibrium processes to a point where it can serve as a
reliable tool for the description of the reaction dynamics from the
initial stage of a heavy ion collision up to the final disassembly
phase.

At relativistic and ultrarelativistic energies the problem of medium
effects on effective masses and cross sections is of paramount
importance.  Hadronic cascade models, which predict compression up to
ten-fold baryon density at the AGS, nevertheless appear to predict
many measured particle yields correctly with the assumption of constant
masses and cross sections.  Double and triple differential cross
sections, as well as improved global event analysis, will provide new
stringent tests for these models.  Theorists have already begun to
question the internal consistency of these models.  More
work along these lines will be needed to identify the crucial
experimental features which may eventually serve as probes for the
properties of hot and dense hadronic matter.

Partonic cascade models are beginning to address the problem of
screening effects which may act as self-consistent cut-off mechanisms
for the infrared divergences of perturbative QCD.  This requires that
the resummation techniques, which have been very successful in
finite-temperature QCD, be extended to strongly interacting matter far
from thermal equilibrium.  The importance of many-body effects
for thermalization of a quark-gluon plasma also deserves further
study.  Progress in this area during the last five years has been
extremely rapid, lending reasonable hope that a consistent description
of very highly energetic nuclear collisions from impact well into the
thermalized plasma phase could be achieved within the framework of
resummed perturbation theory.

There is also a need for definitive Monte Carlo calculations of lattice
gauge theory to determine the precise characteristics of the QCD phase
transition in the presence of physical up, down, and strange quarks.
Questions to be answered include: What is the order of the phase transition?
If it is first order, what is the latent heat? If it is second order, what
is the universality class? What is the interface energy at the phase
boundary and what is the quantitative dependence of the energy
density, entropy, and pressure on temperature? How does the spectrum of
dynamical modes change through the phase transitions?  How are hadron
masses affected below $T_c$?  Given the magnitude of the commitment to
the RHIC project, a more substantial effort of the nuclear theory
community is highly desirable.

The formulation of an effective dynamical theory of hadronization
remains another important challenge.  There exists a wealth of data
from e$^+$e$^-$ and p\=p interactions that can be used to test
hadronization models and to determine the parameters of effective
theories.  The reliability of several proposed quark-gluon plasma
signatures, such as strangeness enhancement and especially
disoriented chiral domain formation, would be significantly enhanced
if the hadronization process were better understood.

In the energy range of the liquid-gas phase transition in nuclear
models, the main challenge is the development of a microscopic,
quantal theory of multifragmentation.  This requires that present
transport models be extended to include the formation and evolution of
density fluctuations.  A solution of this fundamental theoretical
problem would most likely also have applications to relativistic heavy
ion collisions, most notably to hadronization.
\newpage

\section{Weak Interactions}

\subsection{Neutrino Properties and Nuclear Beta Decay}

Although neither Dirac nor Majorana neutrino masses are included in the
minimal standard model, there are strong theoretical arguments suggesting
neutrinos may be massive, and nuclear physics has played a major role in both
direct and indirect tests of this.

The best direct determinations of neutrino masses are the precise
measurements of the end-point region of the tritium $\beta$-decay
spectrum.  Nuclear theorists have contributed to this effort by evaluating
Coulomb corrections to the passage of the outgoing $\beta$-particle through
the molecule and by estimating the effects of atomic and molecular excited
states on the shape of the end-point spectrum.

Neutrino mixing is also expected in most extended models.  The
resulting oscillations can provide information on both
masses and mixing angles.  Furthermore, the effects of modest
mixing can be greatly magnified by matter effects.
Theoretical characterizations of the spectra of solar and
supernova neutrinos provide important yardsticks against which the
effects of oscillations can be measured.
Nuclear theory has made important contributions to the
understanding of signals and backgrounds in neutrino detectors
such as KARMEN and LSND.  This includes modeling the first-forbidden
and quasielastic nuclear responses, and understanding the spallation
products that could complicate certain experiments.

Neutrinoless double beta decay probes the masses and right-handed couplings of
Majorana neutrinos.  Great efforts by nuclear physics have led to agreement on
the size of the matrix elements that govern this process, and thus on the scale
of Majorana masses that current experiments can probe (about 1 eV/$c^2$). The
standard-model process of two-neutrino decay, only recently observed in the
laboratory, provides a very stringent test of nuclear structure theory.
Shell model studies for
nuclei like $^{76}$Ge and $^{82}$Se have successfully predicted rates, while
QRPA calculations in heavier systems have defined the ``reasonable ranges" for
experiment. Recently both Lanczos moment techniques and Monte Carlo methods
have
been successfully used to calculate the nuclear Green's functions that govern
this process.

\subsection{$\beta$-Decay and Unitarity of the CKM Mixing Matrix}

Precision measurements of Fermi $\beta$-decay combined with careful
theoretical analyses of both inner and outer radiative corrections play
a crucial role in testing the unitarity of the Kobayashi-Maskawa matrix.
The error on $V_{ud}$ completely dominates this test, which at the present
time fails at the 2$\sigma$ level.  Reducing theoretical uncertainties is
essential to further progress. Improvements in treating isospin violation
in the nucleus and providing estimates of dispersive effects are important
future challenges for theorists.

New theoretical work has been important to progress in a number
of other $\beta$-decay studies, including the demonstration of large
exchange current effects in axial-charge $\beta$-decay and
the extraction of constraints on scalar and tensor interactions and
right-handed currents.  The origin of the renormalization of $g_A$ in nuclei
and the extraction of $F_P$ from nuclear muon capture and radiative
muon capture remain challenging problems.

\subsection{Weak Lepton-Nucleon and Lepton-Nucleus Scattering}

The measurements of the spin structure function of the proton and neutron
led to various theoretical speculations regarding the polarization of sea
quarks in the nucleon. Spatial as well as spin polarizations of the sea quarks
contribute to the nucleon electroweak currents. Several experiments are
presently being mounted or performed for the purpose of measuring the neutral
weak currents, using either elastic neutrino scattering or parity-violating
elastic electron scattering. Because the photon and the $Z$ boson couple to the
quark flavors with different strengths, comparing the electromagnetic and
neutral weak currents (and assuming that the proton and neutron are an
iso-doublet) allows one to extract the contributions of the individual flavors.
In particular the contribution of $s$ quarks is of special interest because
they
exist only in the quark-antiquark sea.

The interpretation of the expected experimental data has been clarified by
theoretical work carried out in the recent years. At the nucleon
level, one needs to evaluate perturbative
contributions, to model the soft contributions (perhaps through
mesonic loops or the Skyrme model), and to find some reasonable
prescription for matching the two.  It also requires an evaluation of
the hadronic contributions to one-loop electroweak radiative corrections
to tree-level amplitudes that could mimic strange quark effects.
At the nuclear level, one must understand many-body effects such
as contributions from non-nucleonic strange quarks, meson exchange
currents, and electroweak dispersion corrections.

\subsection{Parity Nonconservation}

Nucleon-nucleon scattering and reactions in nuclei provide the only practical
tests of the flavor-conserving nonleptonic weak interaction.  Beautifully
precise measurements in the pp and  few-nucleon systems  and in special light
nuclei such as $^{18}$F and $^{19}$F have shown that the isoscalar
parity-nonconserving interaction has the expected strength, but the isovector
interaction is suppressed relative to expectations.  As the isovector
interaction is thought to be dominated by the weak neutral current, the dynamic
origin of this suppression is of great interest. Theory played a crucial role
in
this physics by providing precise calculations in few-nucleon systems and by
accurately determining transition matrix elements for $^{18}$F and $^{19}$F.
These results have stimulated new work on enhanced operators
associated with strange quarks and on nonperturbative effects
introduced through condensates.

Exploiting the unique beams of epithermal neutrons at the LANSCE
facility, measurements of the longitudinal spin asymmetry have recently
demonstrated parity violation at the level of several percent in multiple
compound-nuclear levels in the same nucleus. The large magnifications
apparent in such measurements are understood in terms of the chaotic, quantum
behavior of the nucleus. Theorists have made great progress in understanding
the observed enhancements in terms of the underlying parity-violating
nucleon-nucleon interaction using methods of statistical spectroscopy,
where the relevant observables are expressed as averages over collections
of states. However, in one case ($^{232}$Th) the observed signs of the
mixing were found to be highly nonstatistical, posing an intriguing
problem to theorists.

It is now widely appreciated that precise atomic PNC measurements can provide
information on new physics complementary to collider data.  This has stimulated
experimental efforts to measure PNC in isotopic chains, where one can factor
out
the complicating effects of atomic physics.  Theoretical studies indicate that
changes in the neutron distributions, which dominate the weak charge, could
complicate the interpretation of these experiments if they achieve their
anticipated 0.1\% accuracy. Progress is therefore likely to depend on improved
calculations and theoretical understanding of neutron distributions and radii.

\subsection{T- and CP-Nonconservation}

The precision of atomic electric dipole moment measurements has reached
that of neutron electric dipole moment experiments and will soon far
surpass it, posing new challenges to nuclear theory.
For many sources of CP nonconservation (CPNC), such as a
CP violating $\theta$-term in the QCD Lagrangian, the atomic electric
dipole moment arises predominantly from CPNC interactions within the
nucleus, making the quality of future CPNC limits dependent on our
ability to calculate the CPNC forces between nucleons that arise from
the underlying Lagrangian and to evaluate the resulting nuclear
polarizations.

Important constraints on CPNC have also been obtained from correlation
studies in $\beta$-decay, helped by theoretical estimates of the final-state
effects.  Constraints on exotic T-odd, P-even nuclear forces have been
obtained from calculations of nuclear electric dipole moments induced by
weak radiative corrections, and it has been recently argued that neutron
transmission experiments similar to those discussed in the section on parity
nonconservation could reach similar sensitivities.

\subsection{Challenges and Opportunities}

Signatures for new physics beyond the standard model may well be first
discovered as a subtle symmetry violation at low energies.
It is extraordinary how many of the crucial tests, such as
small neutrino masses, electric dipole moments of atomic nuclei, and
family-number violating muon decays, are now reaching sensitivity levels
characteristic of expected new physics.  Nuclear theorists will continue
to have an important part in connecting the observed phenomena, or
bounds on such phenomena, with the underlying fundamental mechanisms.
\newpage

\section{Nuclear Astrophysics}

Together, nuclear physics and astrophysics have greatly enhanced our
understanding of the universe.  A large variety of phenomena in the universe,
such as primordial and galactic nucleosynthesis, stellar evolution, neutrino
luminosities, supernovae and neutron stars, require quantitative understanding
of nuclear processes.  The synergy between these two fields will grow further
when the proposed radioactive ion beam facilities provide new information on
nuclei far from stability and when the QCD structure of nuclear and hadronic
matter is better understood via the CEBAF and RHIC programs.

\subsection {The Solar Neutrino Problem}

New results from the Kamiokande, SAGE, and GALLEX observatories have
revealed a pattern of solar neutrino fluxes which is difficult to reconcile
with the possible changes in the standard solar model. At the same time,
helioseismology provides new constraints on permissible solar model
variations. The result is an emerging consensus that the solar
neutrino problem most likely involves physics beyond the standard model.
The discovery of the MSW mechanism, a solar enhancement of flavor
violation engendered by adiabatic level crossings, provides an elegant
explanation for the experimental results.

Nuclear theorists played a prominent role in developing many of the
analytical and numerical tools for treating the propagation of neutrinos
through stellar medium. They have also been involved with proposing new
experiments like Borexino and $^{127}$I, and with the efforts
to reduce detector cross section uncertainties for $^{37}$Cl, SNO and
SAGE/GALLEX observatories.

The primary remaining nuclear uncertainty in the solar neutrino problem is
the S-factor for $^7$Be(p,$\gamma$)$^{8}$B reaction. The two available
low-energy data sets disagree by $\approx 25\%$.
Noticeable progress has been achieved in the theoretical description of the
Coulomb dissociation process which can be used to measure this rate using
radioactive $^8$B beams. Ab initio calculations
of the pp-chain reactions are now within reach of quantum Monte
Carlo methods. Initial studies have focused on the weak radiative capture
of protons by $^3$He, the source of the most energetic solar neutrinos.

\subsection {Nucleosynthesis}

Big-bang nucleosynthesis is one of the three cornerstones of modern
cosmology. Motivated by the possibility of violent
confinement/deconfinement or electroweak phase transitions, nuclear
theorists and experimentalists
have explored the consequences of inhomogeneities on big-bang
nucleosynthesis. Careful work has shown
that nucleosynthesis places stringent constraints on such scenarios.

The observed abundances of the presupernova elements up to zinc can
now be well reproduced within models of the galactic chemical evolution.
Nevertheless some astrophysically important nuclear cross sections are
still insufficiently known. Most important is the $^{12}$C($\alpha,
\gamma$)$^{16}$O reaction rate which plays a decisive role in both
nucleosynthesis and the evolution of massive stars. The theory-stimulated
experimental determination of the electric dipole part of this rate was a major
milestone of the field in recent years. However, while microscopic cluster
models predict that the $^{12}$C($\alpha,
\gamma$)$^{16}$O electric quadrupole S-factor is comparable to the dipole,
experimental confirmation is still missing.

The new radioactive ion beam facilities will allow measurements of
important reactions that facilitate the leakage
of material from the hot CNO cycle, which subsequently leads to
nucleosynthesis of elements up to $^{56}$Ni and
beyond by rapid proton capture. The theoretical challenge here is to
extrapolate the data to the astrophysically important
low energies. The predicted enhancement of the low-energy fusion cross
section due to partial screening of the nuclear charges by electrons in
the target has been observed. Current theoretical modeling underestimates
the screening effects for most of the reactions experimentally studied. It
is critically important that this discrepancy be resolved, in order to
make use of the remarkable efforts currently being made to push laboratory
cross-section measurements to still lower energies.

Core collapse supernovae are the major engines driving galactic chemical
evolution, ejecting both the hydrostatic burning products and the many
less abundant species made by the explosion itself. Two important
advances have recently occurred in this area. Network simulations have
demonstrated that r-process nucleosynthesis takes place in the hot bubble
outside the neutrinosphere of a core-collapse supernova, where neutron
rich material driven by neutrino wind mixes with seed nuclei produced by
$\alpha$-burning. For the first
time the observed pattern and yield of the synthesized elements can be
understood. Second, careful modeling of the
inelastic neutral current interactions of neutrinos with nuclei in the
mantle of the supernova revealed a new nucleosynthesis mechanism, the
neutrino process. New elements are produced by spallation
following excitation of giant resonances by inelastic neutrino scattering.
The process resolves long-standing
puzzles such as the origin of $^{19}$F and the inability of cosmic ray
spallation mechanisms to produce the correct $^{11}$B/$^{10}$B ratio.

Simulations of the r-process depend on properties and masses of nuclei near the
neutron drip line. Hartree-Fock-Bogoliubov
calculations predict a closing of shell gaps for closed neutron shell
nuclei near the drip line that, when incorporated
into r-process simulations, helps to fill the abundance valley below the
mass peak at $A=130-135$, in agreement with observation.

Properties of unstable proton-rich nuclei are important to the
understanding of the rapid proton capture process
which powers the (Type I) outbursts of x-ray bursters. Their understanding
also provides a motivation for continued experimental and theoretical
studies of nuclei far from stability.

\subsection {Supernovae}

An important recent advance has been the development of two-dimensional
hydrodynamical codes for simulating stellar collapse.
These calculations find convective cells forming above the neutrinosphere
which sweep colder matter to smaller radii, where
it can be more effectively heated by neutrinos. Hot material is carried to
the shock by buoyancy without having to pay a large
gravitational penalty.  Convection appears to increase the deposition of
neutrino energy into the shock to the point where the delayed explosion
mechanism succeeds. This could be the long-sought solution of the supernova
problem. The important challenge is to demonstrate
that this success persists when all of the physics is fully modeled,
including a state-of-art treatment of neutrino diffusion, and that
ejection of synthesized nuclei occurs. Screening effects for weak
interactions in a relativistic plasma, which can
substantially increase neutrino mean free paths in matter at densities
above 10$^{12}$ g/cm$^3$ are being studied along with coupling of plasmon
excitations to neutrinos.

An interesting result, independent of the details of the explosion, is that
the temperatures of the heavy flavor neutrinos
are about 8 MeV while those of electron flavor neutrinos are about 4 MeV.
It has been recently emphasized that the cosmologically interesting
heavy-flavor neutrinos may undergo an adiabatic MSW crossing outside
the neutrinosphere. Matter-enhanced mixing between $\nu_e$ and either
$\nu_{\mu}$ or $\nu_{\tau}$
occurring anywhere above the neutrinosphere will cause
a characteristic temperature inversion, and a distinctive signal in
terrestrial detectors sensitive to $\nu_e$'s and $\bar{\nu}_e$'s.

The Gamow-Teller (GT) strength of several fp-shell nuclei, important for
the electron capture process in the early collapse
stage, has been determined experimentally from (n,p) charge exchange
reactions. The shell model Monte-Carlo and Lanczos
techniques can successfully reproduce the observed strength. The
Monte-Carlo method also allows studies of nuclei at
finite temperatures and suggest that the GT strength is roughly constant
for $T \leq 2$ MeV. It is found that, in even-even
fp-shell nuclei, the pairing between like particles vanishes at
temperatures around 1 MeV. The consequences of this phase
transition for astrophysical scenarios have still to be explored. Neutral
current decay of nuclear excited states, by pair neutrino emission,
during the collapse, have also been calculated.

\subsection {Neutron Stars}

Ab initio calculations may be possible for neutron stars due to their
quasi-static condition, however, they pose many new challenges.
The predictions of conventional nuclear many body theory are in accord with
the available data on masses, surface red shifts,
rotational periods and temperatures of neutron stars. The maximum density
of matter in the core of commonly observed $1.4 M_{\odot}$ stars
is estimated to be about four times that in nuclei. Interestingly, nuclear
forces seem to give rather small contributions
to the energy of matter at such densities, as compared to the Fermi kinetic
energy; however they enhance the pressure significantly.
In heavier stars, near the maximum mass limit, matter is at much higher
density and nuclear forces give large contributions with
significant uncertainties.

Substantial progress has been made in the theoretical predictions of the
structure of mater in the inner crust, where nuclear
matter containing neutrons and protons coexists with pure neutron matter.
As the density increases to about half nuclear density the
structure of matter changes from that of having drops to rods to sheets of
nuclear matter in neutron matter. The pure neutron matter
in the crust is predicted to be a superfluid whose angular momentum is
carried by vortices spatially pinned by the nuclear matter
drops and rods. The dynamics of these vortices, which raise theoretical issues
also of interest in condensed matter physics, provide one plausible
explanation for the observed glitches in neutron stars.

The possible occurrence of new, exotic phenomena in the core of neutron
stars is of great interest. It has been realized recently that
drops and rods of quark matter may coexist with hadronic
matter over an extended region within neutron stars. The question of
pion condensation in neutron star matter is still open. Recent
speculations regarding kaon condensation in the core, based on
effective chiral Lagrangians, have triggered a great deal of interest in
kaon-nucleus interactions.

\subsection {Exotic Particles}

Weakly interacting massive particles (WIMPS) are a leading dark matter
candidate. Experiments exploiting existing and new detectors (e.g.,
cryogenic Si detectors) have placed important limits on WIMPS.
Nuclear theorists have estimated the detector cross sections, including
form factors that are important at the expected large momentum transfers.

Modeling of supernovae, and the subsequent understanding of their cooling
curves, has provided constraints on the properties
of new particles or interactions that could alter that cooling. Some of the
important constraints include those on axions, Majorons,
neutrino magnetic moments, and neutrino Dirac masses.

\subsection{Challenges and Outlook}

Solving the solar neutrino problem, successfully modeling the supernova
explosion mechanism, and understanding supernova
heavy element nucleosynthesis are three immediate problems in stellar
physics. They contain new, interesting aspects of
neutrino propagation in dense matter. If the resolution of the first involves
new neutrino physics, the field is positioned to contribute to the
construction of a new standard model. The latter two problems are intimately
coupled, since the dynamics of the supernova explosion and the fossil
record of that explosion in the synthesized nuclei must
be understood within a single model.

Calculations of high density neutron star matter is still an active field
with many uncertainties. Better understanding, based on QCD and constrained
by available data, of the two- and three-hadron interactions is needed.
It is also necessary to determine the
density at which the transition from cold hadronic to quark matter occurs.
Extracting information on the properties of neutron star matter from the
ongoing observations of neutron stars is also a challenging problem.

In the longer term, it is apparent that the nature of astrophysics and
astronomy is rapidly changing.  The explosion of new instrumentation is
providing detailed information on the universe.  Increasingly, the
interpretation of these new data depends on our understanding of the
underlying atomic and nuclear microphysics.  Thus our field's partnership
with astrophysics will be increasingly important in explaining the
phenomena we discover around us.
\newpage

\section{Connections of Nuclear Theory to Other Areas}

\subsection{Atomic Clusters}

Nuclear physics has been seminal for the development of
atomic cluster physics.  Since the discovery of spherical magic
numbers in the mid 1980's, theoretical ideas from nuclear physics
have made several important contributions.
Unlike nuclei, atomic clusters can be made with thousands of
particles, and for such large systems a new shell phenomenon,
the supershell, was predicted by nuclear theorists.
This has seen spectacular confirmation in experiments on sodium
clusters, which show shells extending up to the
thousands of atoms, and a supershell minimum near $N=1000$.

\begin{figure}[h]
\def\epsfsize#1#2{0.7#1}
\centerline{\epsfbox{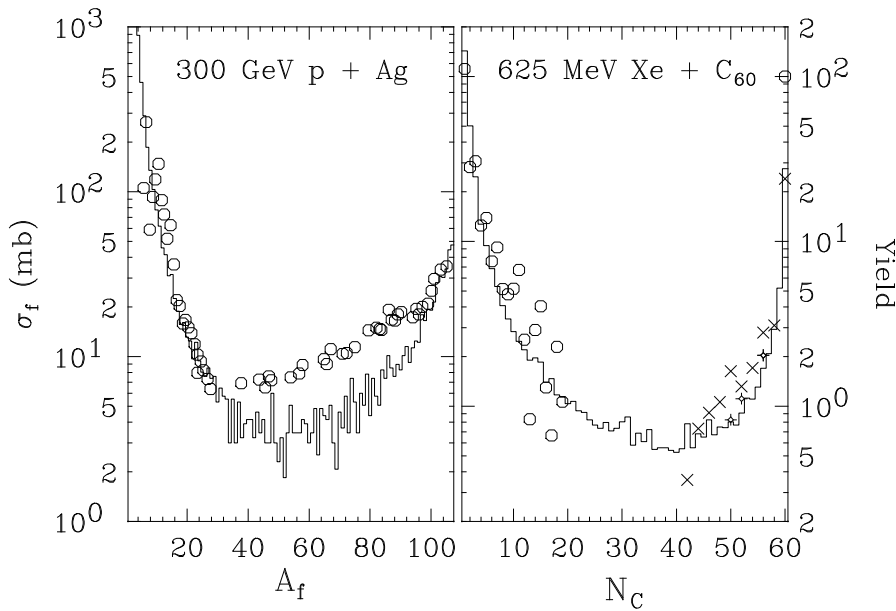}}
\caption{
Fragmentation data compared with the percolation model.  On
the left is shown the probability of forming fragments with
atomic number $A$ when silver nuclei are bombarded by
high energy protons.  The right part of the figure
shows the fragmentation of C$_{60}$ clusters by a Xe beam.
}
\end{figure}

The well-known giant dipole resonance in nuclear physics has a close
analog in the Mie resonance of simple metal clusters.  Many of
the properties of the Mie resonance were anticipated from the
nuclear example:  deformed splitting, thermal broadening and the
existence of a collective mode in C$_{60}$.

The nuclear physics stimulus has also led to new directions in
cluster reaction studies.  The fission of charged clusters
is one example.  Another is the fragmentation of clusters
by a high-energy probe.  The data show intriguing similarities
to nuclear fragmentation, and simple theory with the percolation
model gives similar rough descriptions of the fragmentation
yields. The comparison is shown in Fig. 6.

Rare gas clusters, especially $^3$He and $^4$He clusters, are even more
like nuclei in that they are dominated by short range interactions.
Nuclear theorists have pioneered the study of quantum helium liquid
droplets by exact quantum Monte Carlo methods  and have continued
to provide leadership in the current research program on impurity
scattering and laser spectroscopy in both physics and chemistry.

\subsection{Mesoscopic Physics}

Many ideas from nuclear physics have been applied to the
study of mesoscopic condensed matter systems.  In particular,
phenomena that involve the discreteness of electron single-
particle levels often have analogies in compound nucleus
theory.  One example is the conductance fluctuations in
small wires and quantum dots. The theory of these fluctuations
for diffusion-limited wires was developed by applying
theory developed in nuclear physics, namely random matrix
models of spectra and precompound reaction theory.
Theory also explains the fluctuations
in the ballistic electron regime with a similar approach to
that used to describe  Ericson fluctuations of nuclear reactions.
Shell physics and its semiclassical description has been
found to be related to the phenomenon of persistent currents in
mesoscopic rings. The magnitude of theses currents depends on an
interplay between the regular spectra of the shell model and chaotic
spectra of disordered systems.  Finally, the quantum dot
provides a close analogy to the compound nucleus.  Its
spectra with isolated resonances gives a new demonstration
of the well-known Porter-Thomas fluctuations.
Multiple scattering theory has also been an extremely useful tool
in understanding the many-electron states in nanoscale condensed
matter structures, such as quantum corrals.

\subsection{Spin Systems}

The close relationship between between lattice field theory and problems
in statistical mechanics provides fertile ground for interdisciplinary
contributions. One example is the contributions nuclear theorists have
made in recent years to the study of spin systems in condensed
matter physics. In classical spin systems, the phase structure of
the planar x-y model on a two-dimensional triangular lattice was determined
and led to the discovery of a new class of multicritical points.
A particularly important quantum spin system is the spin
$1/2$ two-dimensional Heisenberg model, relevant to high temperature
superconductors. For this system, several new approaches were
introduced, including a loop cluster algorithm
utilizing an improved estimator which has allowed the most accurate Monte
Carlo calculation to date of the low energy parameters which are in agreement
with the experimental data for precursor insulators of high $T_c$
superconductors.

\subsection{Molecular Physics}

While algebraic methods were initially developed to make the study of finite,
strongly--interacting many--body nuclear  systems tractable, the formal
techniques are broad in scope and have application to other disciplines. One
of the areas in which such nuclear physics techniques have had a particularly
large impact is molecular physics, in particular, the use of algebraic theory
as a way to describe molecular interactions. By converting
the differential Schr\"{o}dinger equations of quantum mechanics into
algebraic equations, it is now possible to attack problems previously deemed
intractable, especially for strongly anharmonic molecules.

Some such problems
include the study of  intramolecular relaxation in large molecules,
in particular how energy is shared between the many degrees of freedom of a
complex molecule; the study of the polymerization process, in particular how
large molecules join to form dimers, trimers, etc., and the role of finite
size effects in polymer chains. One can now also compute the complete
thermodynamics of complex molecules, including the normalized density of
states. Because the computational problem grows only linearly with the
number of bonds, it is  ideally suited for the study of large molecules and
polymers. The understanding of molecular geometries and bond angles as
functions  of temperatures, as well as the physics of phase transitions,
can be treated with  nuclear mean field methods.

Nuclear physics techniques  have also had an impact on more formal aspects of
point groups. One can now incorporate discrete symmetries directly
into the Hamiltonian, so that the study of large molecules  such as C$_{60}$,
with icosahedral symmetry, is possible.  Further, since anharmonicities can
be introduced from the outset,  one can study highly excited states of
molecules, which is especially  crucial for vibrations of CH and OH
chromophores.

The formalism used in nuclear physics which determined the
scattering matrix for proton scattering from nuclei to all orders in the
eikonal approximation has been applied to electron scattering
from molecules with great success.
\newpage

\section{Computational Challenges}

A recurring theme in the challenges and opportunities for the future is the
exploitation of emerging computer resources to make major advances in the
solution of previously intractable many-body and field theory problems. From
a qualitative advance in the ability to solve the interacting shell
model and the development of exact solutions for the response
functions of light nuclei to quantitative lattice calculations of
hadronic observables central to understanding the physics on frontier
accelerators, advanced computation provides unprecedented opportunities
for fundamental developments in nuclear science.   In addition, maintaining
our traditional leadership role in computational science has important
societal benefits, ranging from the  education of students
and postdocs in computational science, which is essential to the
scientific infrastructure, to playing a leadership role in
the development of high performance computer technology in this country.

Nuclear science needs to provide  adequate access to state-of-the
art computer technology at every level. At the highest end, it should
take part in pursuing Teraflops-scale computation in this country,
contributing to an effort
commensurate with the investment being made for example  by Italy and
Japan. It should also take the lead in providing the optimal balance of
local versus centralized facilities, and in pursuing the exploitation
of cost-effective workstation farms and symmetric multiprocessors.

An indication of the scale of resources required to meet the immediate
challenges can be seen by looking at Table 1, giving a list of examples
for the scale of resources used in current efforts in some of
the areas in which we have cited opportunities. For convenience,
resources are converted to Gigaflops-years. One Gigaflops-year represents
the dedicated use of a computer at a sustained rate of one billion
floating point operations per second (``flops'') for the period of one year,
or the completion of $3 \times 10^{16}$ floating point operations.  For
reference, the total resources allocated to nuclear theory at NERSC by the
DOE corresponds to about 3 Gflops-years.

\begin{table}[b]
\begin{center}
\begin{tabular}{|l|c|}
\hline
Research & Gflops-years \\
\hline
Lattice QCD & 10 \\
Monte Carlo Shell Model	& 0.5 \\
Conventional Shell Model & 1 \\
Mean-field Theory & 0.5 \\
Variational Monte Carlo	& 0.5 \\
Green's Function Monte Carlo & 1 \\
\hline
\end{tabular}
\caption{Currently used resources (in Gflops) in some areas of
computationally intensive nuclear physics}
\end{center}
\end{table}

To fully exploit emerging opportunities, these resources must grow
significantly. For example, in Europe, the dedicated APE QCD machines now
provide up to 100 Gflops, and significant new resources would be required
in the U. S. to remain competitive. Similarly, substantial increases in the
resources devoted to Monte Carlo calculations would be required to accomplish
the goals described in this report.

\section{Appendix: Town Meeting Program}

\subsubsection*{Sunday, January 29:}
\medskip

\noindent 8:30am -- 9:40am {\it Reports from other Town Meetings}
(Chairman: J. Friar)
\medskip

\begin{tabular}{lll}
&8:30am  &Welcome (Berndt M\"uller) \\
&8:35am &Technical remarks (Robert Wiringa) \\
&8:40am &Nuclear structure, low energy reactions, etc.
(Witold Nazarewicz) \\
&8:55am &Electromagnetic probes (Xiangdong Ji) \\
&9:10am &High energy heavy ions (Xin-nian Wang) \\
&9:25am &Hadronic probes (Mikkel Johnson) \\
\end{tabular}
\medskip

\noindent 10:00am -- 12:00 {\it Nuclear Structure}
(Chairman: F. Iachello)
\medskip

\begin{tabular}{lll}
&10:00am  &Structure of normal and exotic nuclei (Joseph Ginocchio,
Los Alamos) \\
&10:15am &Few-body systems and nuclear matter (Rocco Schiavilla,
CEBAF/ODU) \\
&10:30am &Exact shell model calculations (David Dean, Caltech) \\
&10:45am &Relativistic many-body theory (Brian Serot, Indiana)  \\
&11:00am &Short presentations (Brown, Zelevinsky, Carlson, Wiringa,
Dickhoff)\\
&11:25am &Open discussion \\
\end{tabular}
\medskip

\noindent 1:30pm -- 4:00pm  {\it Quantum Chromodynamics}
(Chairman: G. Bertsch)
\medskip

\begin{tabular}{lll}
&1:30pm &QCD and hadron structure (David Kaplan, Seattle) \\
&1:45pm &Lattice gauge theory (John Negele, MIT) \\
&2:00pm &QCD vacuum and sum rules (Edward Shuryak, Stony Brook) \\
&2:15pm &QCD light cone approach (Robert Perry, Ohio State U.) \\
&2:30pm &Quark gluon plasma (Joseph Kapusta, Minnesota) \\
&2:45pm &Short presentations (Roberts, Qiu, Strikman, Ji, Banerjee)\\
&3:10pm &Open discussion \\
\end{tabular}
\medskip

\noindent 4:30pm -- 6:30pm {\it Nuclear astrophysics and other
topics} (Chairman: V. Pandharipande)
\medskip

\begin{tabular}{lll}
&4:30pm &Nuclear astrophysics (Karl-Heinz Langanke, Caltech) \\
&4:45pm &Weak Interactions (Michael Musolf, CEBAF/ODU) \\
&5:00pm &Nuclear transport theory, (J\o rgen Randrup, LBL) \\
&5:15pm &Applications to other fields (Aurel Bulgac, Seattle) \\
&5:30pm &Short presentations (Olinto, Kim, Bauer, Elster)\\
&5:50pm &Open discussion \\
\medskip

&6:30pm &{\it Adjourn} \\
\end{tabular}
\bigskip

\noindent {\bf Monday, January 30:}
\medskip

\noindent 8:30am -- 10:00am {\it Community and infrastructure
issues} (Chairman: J. Negele)
\medskip

\begin{tabular}{lll}
&8:30am &Outlook at DOE, manpower and jobs (Joseph McGrory, DOE) \\
&8:45am &Outlook at NSF, manpower and jobs (Bradley Keister, NSF) \\
&9:00am &Role of the Nuclear Theory Institute (Wick Haxton, Seattle) \\
&9:15am &Short presentations and open discussion \\
\end{tabular}
\medskip

\noindent 10:30am -- 12:00  {\it General discussion of priorities}
(Chairman, B. M\"uller)

\end{document}